\documentclass[11pt]{article}

\usepackage{amsfonts}
\usepackage[centertags]{amsmath}
\usepackage{amssymb}
\usepackage{cite}
\usepackage{psfrag}
\usepackage{graphicx}
\usepackage{bbm,bm}
\usepackage{epsfig, multicol}
\usepackage{hyperref}

\allowdisplaybreaks
\begin{document}

\topmargin 0pt
\oddsidemargin 0mm
\def\be{\begin{equation}}
\def\ee{\end{equation}}
\def\bea{\begin{eqnarray}}
\def\eea{\end{eqnarray}}
\def\ba{\begin{array}}
\def\ea{\end{array}}
\def\ben{\begin{enumerate}}
\def\een{\end{enumerate}}
\def\nab{\bigtriangledown}
\def\tpi{\tilde\Phi}
\def\nnu{\nonumber}
\newcommand{\eqn}[1]{(\ref{#1})}

\newcommand{\half}{{\frac{1}{2}}}
\newcommand{\vs}[1]{\vspace{#1 mm}}
\newcommand{\dsl}{\pa \kern-0.5em /} 
\def\a{\alpha}
\def\b{\beta}
\def\g{\gamma}\def\G{\Gamma}
\def\d{\delta}\def\D{\Delta}
\def\ep{\epsilon}
\def\et{\eta}
\def\z{\zeta}
\def\t{\theta}\def\T{\Theta}
\def\l{\lambda}\def\L{\Lambda}
\def\m{\mu}
\def\f{\phi}\def\F{\Phi}
\def\n{\nu}
\def\p{\psi}\def\P{\Psi}
\def\r{\rho}
\def\s{\sigma}\def\S{\Sigma}
\def\ta{\tau}
\def\x{\chi}
\def\o{\omega}\def\O{\Omega}
\def\k{\kappa}
\def\pa {\partial}
\def\ov{\over}
\def\nn{\nonumber\\}
\def\ud{\underline}
\def\qq{$Q{\bar Q}$}
\begin{flushright}
%
\end{flushright}
\begin{center}
{\Large{\bf Wilson loop calculation in QGP using\\ non-supersymmetric AdS/CFT}}

\vs{10}

{Somdeb Chakraborty$^a$\footnote{E-mail: somdeb.ch.12@gmail.com}, Kuntal Nayek$^b$\footnote{E-mail: kuntal.nayek@saha.ac.in} 
and Shibaji Roy$^b$\footnote{E-mail: shibaji.roy@saha.ac.in}}

\vs{4}

$^a$ {\it Department of Physics, City College\\ 
102/1, Raja Rammohan Sarani, Calcutta 700009, India}

\vs{4}

$^b$ {\it Saha Institute of Nuclear Physics\\
1/AF Bidhannagar, Calcutta 700064, India}

\vs{4}

{\rm and}

\vs{4}

{\it Homi Bhabha National Institute\\
Training School Complex, Anushakti Nagar, Mumbai 400085, India}
\end{center}

\vs{10}

\begin{abstract}
Previously, two of the present authors obtained the decoupling limit and the corresponding throat geometry of 
non-supersymmetric D3 brane solution of type IIB string theory. In analogy with the supersymmetric case, it 
describes the gravity dual of a non-supersymmetric gauge theory with QCD-like properties such as running coupling
and confinement (or mass gap) in certain range of its parameters. In this paper, we consider a `black' 
version of the non-supersymmetric D3 brane solution in the decoupling limit and use this gravity background to 
holographically compute the expectation value of a time-like Wilson loop which, in turn, is related to the potential 
of a heavy quark-antiquark pair. By boosting the gravity solution along one of the brane directions and placing 
the pair at an arbitrary orientation with this direction, we numerically obtain the variation of the screening 
length as well as the potential with velocity, its orientation with respect to the direction of motion and 
other parameters of the theory. Remarkably enough, our results are in qualitative agreement with those obtained 
holographically in supersymmetric gauge theories indicating that these features are quite robust and universal 
as they are insensitive to the presence of any supersymmetry in the theory. The physical interpretations of 
the variations with respect to the other parameters of the theory, not observed in supersymmetric theory, have also been 
given.            
\end{abstract}


\noindent{\it 1. Introduction} : The AdS/CFT correspondence \cite{Maldacena:1997re, Witten:1998qj, Gubser:1998bc} 
is a conjectured equivalence between two theories - one, 
a string theory in a particular background and, therefore, includes gravity, while the other is a quantum field theory 
without gravity. It is a holographic correspondence in the sense that a field theory in certain space-time dimensions is 
related to string theory living in a background space-time of one dimension higher. It is also a strong/weak duality - 
when the field theory is strongly coupled, the dual string theory is weakly coupled, i.e., given by supergravity and 
{\it vice-versa}. The correspondence has since been generalized to encompass a wider variety of gauge theories with different 
gravity duals and is now more aptly called the gauge-gravity duality \cite{Aharony:1999ti}. The duality has proved to be an extremely powerful 
tool that allows us to gain valuable insights into the behavior of strongly coupled field theories (by mapping the system 
to a suitable holographic dual gravity theory)  which are otherwise not accessible via the standard perturbative formalism. 
In the original proposal, both sides of the duality are required to respect supersymmetry and conformal symmetry. In 
particular, on the string theory side, one takes a large number $(N)$ of coincident BPS D3 branes of type IIB string theory 
and looks at the decoupled geometry (throat of the D3 brane) near the branes which is a maximally symmetric 
AdS$_5$ ($\times$ S$^5$) space and relates it with $(3+1)$-dimensional ${\cal N}$=4, SU($N$) super Yang-Mills (SYM) theory 
at large $N$, living on the boundary of AdS space. 

It is well-known that when heavy ions like gold or lead, moving in opposite directions, collide head-on at ultra-relativistic 
energies, they produce a fluid-like state of matter made up of strongly interacting quarks and gluons better known as 
quark-gluon plasma (QGP) (see \cite{Shuryak:2014zxa} for a recent review). Many of the properties of QGP like thermalization, 
chiral symmetry breaking, deconfinement, 
etc. have been studied in the past in RHIC and is also being currently studied extensively in LHC (For example, see
this comprehensive review \cite{CasalderreySolana:2011us} and the references therein.). 
The theoretical frameworks 
at our disposal to explore this strongly coupled plasma are primarily lattice QCD and the AdS/CFT duality. Unlike lattice field 
theory, the AdS/CFT correspondence is tailor-made for studying the real-time dynamics that is of interest in many cases. In 
particular, the AdS/CFT formalism has been extensively used to study, among other properties, the screening length of a heavy 
quark-antiquark ($Q{\bar Q}$) pair as well as its potential \cite{Liu:2006ug, Liu:2006nn, Liu:2006he}. In the holographic picture, 
one takes a stack of BPS D3 branes in 
the decoupling limit which is the gravity dual of $D=4$, ${\cal N}$=4 SYM theory at large $N$ and introduces a probe fundamental string 
in this background. To examine the effect of velocity on the $Q{\bar Q}$ pair, one boosts the gravity solution along a brane 
direction and to incorporate the effect of a non-zero temperature, one considers a `black' brane. However, QCD being neither 
supersymmetric nor conformal, one may question whether BPS D3 brane is the appropriate framework to study the properties of QGP 
from a holographic perspective. This motivates us to consider non-supersymmetric gravity solutions that can potentially be used 
to model real-world QCD more faithfully. While the literature abounds with works addressing various observables related to 
supersymmetric gauge theories using the holographic correspondence \cite{CasalderreySolana:2011us}, our purpose here is to examine 
whether the lack of 
supersymmetry has any significant effect on the qualitative features of heavy quark observables. Non-supersymmetric backgrounds 
representing the holographic dual of QCD-like theories have been constructed earlier from string theory 
\cite{Witten:1998zw, Sakai:2004cn, Klebanov:1998yya} 
as well as from 
phenomenological point of view \cite{Polchinski:2001tt, Erlich:2005qh, Andreev:2006ct} to study various aspects of QCD using the 
gauge/gravity duality. Type II string theories are known 
to admit BPS as well as non-supersymmetric (non-susy) D$p$ brane solutions \cite{Zhou:1999nm, Brax:2000cf, Lu:2004ms}. If, like 
BPS branes, the non-susy branes also have 
a decoupling limit, they will naturally represent the gravity dual of a non-supersymmetric gauge theory like QCD. Indeed, in our 
recent works, we have shown, by studying graviton scattering in the background of non-susy D$p$ branes, that bulk gravity gets 
decoupled on the brane very similar to the case of BPS branes \cite{Nayek:2015tta}. We also worked out the details of the decoupling limit for non-susy 
D3 branes and obtained the throat geometry \cite{Nayek:2016hsi}. This geometry, under a suitable coordinate transformation, has been shown to match 
with the Constable-Myers solution having many interesting properties like confinement and running coupling constant similar 
to QCD \cite{Constable:1999ch}. 
In this paper, we consider a `black' version of this solution which corresponds to the holographic dual of a non-susy gauge theory 
at finite temperature \cite{Bai:2006vv, Lu:2007bu}. We intend to study the screening length and the potential of a heavy $Q{\bar Q}$ pair (equivalently, called 
a dipole) moving through a hot plasma whose gravity dual is the non-susy background just mentioned. Our computation is similar in 
spirit to the one performed by Liu, Rajagopal and Wiedemann (LRW) \cite{Liu:2006ug, Liu:2006he} for the $D=4$, ${\cal N}$ = 4 SYM plasma. 
We introduce a fundamental string 
as a probe whose end points (representing the $Q{\bar Q}$ pair), separated by a distance $\ell$, lie on the boundary in the 
$x^1$-$x^3$ plane and makes an angle $\theta$ with the $x^3$ direction, along which we have given the background a boost. Following 
LRW and the holographic dictionary, we compute 
the thermal expectation value of the time-like Wilson loop\footnote{See \cite{Maldacena:1998im, Rey:1998ik, Brandhuber:1998er} 
for some early computations of Wilson loop in AdS/CFT.} 
$\langle W^F({\cal C})\rangle$ in the fundamental representation by 
calculating the minimal world-sheet area swept out by the open string with a boundary which coincides with the loop ${\cal C}$. 
The precise relation between them is 
$\langle W^F({\cal C})\rangle = {\rm exp}[i\mathcal{S}({\cal C})] = {\rm exp}[i E({\cal C}) {\cal T}]$, where $\mathcal{S}({\cal C})$ 
is the action (finite) for the extremal world-sheet and for time-like Wilson loop, $\mathcal{S}({\cal C})$ is proportional to time 
${\cal T}$, and, therefore, $E({\cal C})$ represents the $Q{\bar Q}$ potential. Typically, this potential suffers from a divergence 
which can be cured by subtracting out the self-energy of the free quark and antiquark. In the process, we also compute the 
separation length of the $Q{\bar Q}$ pair or the dipole length whose maximum value gives the screening length. In the absence of 
analytical expressions, we obtain numerically the variations of the screening length 
as well as the potential with the velocity, orientation of the dipole with respect to the direction of the moving plasma and other 
parameters of the theory. Comparison of the obtained results with the supersymmetric LRW \cite{Liu:2006he} counterparts leads us to the interesting 
observation that all the results exhibit qualitatively similar pattern irrespective of the presence of supersymmetry. That supersymmetry 
need not be an essential ingredient for these features to exist, is an important step towards a better understanding of QGP since the 
plasma generated in the collider experiments is not itself supersymmetric. The other parameters will be shown to be related to the
temperature and the coupling in the boundary theory.   

This paper is organized as follows. In section 2, we review the non-susy D3 brane solution and its decoupling limit. Section 3 is 
devoted to the calculation of the thermal expectation value of the time-like Wilson loop from which we extract the formal 
expressions for the $Q{\bar Q}$ separation length and the $Q{\bar Q}$ potential. This is followed by the numerical results and 
their discussion in section 4. Finally, we conclude in section 5.

\vspace{.5cm}

\noindent{{\it 2. The `black' non-susy D3 brane and the decoupling limit}} : The non-supersymmetric D3 brane solution and 
its decoupling limit has been discussed in \cite{Nayek:2016hsi}. Here, we consider a `black' version of this solution. We take the non-susy 
D$p$ brane solution, anisotropic in $t$ as well as one of the brane directions $x^1$, given in eqs.(4) and (5) of ref.\cite{Lu:2007bu}. 
For D3 brane, we put $p=3$ and make it anisotropic only in $t$ direction by setting $\d_2=\d_0$ and ${\bar \d} = (3/4) \d_2$. 
We further put $\d_1+2\d_2 = \d$ which enables us to eliminate $\d_2$ as an independent parameter. The resulting solution takes 
the form\footnote{Note that the metric in the solution does not have the full Poincare symmetry ISO(1, 3) in the brane world-volume
directions, rather it is broken
to R $\times$ ISO(3) and that is the reason we call it `black' non-susy D3 brane solution. However, we put black
in inverted comma because this solution does not have a regular horizon like true black brane, rather, it has a singular horizon. 
But still one can define a temperature for this solution as we mention later.}
\bea\label{isolution}
ds^2 & = & F(r)^{-\half} \left(\frac{H(r)}{\tilde{H}(r)}\right)^{-\frac{\d}{4}-\frac{3\d_1}{8}}\left[\left(\frac{H(r)}{\tilde{H}(r)}\right)^\d (-dt^2) 
+ \sum_{i=1}^3 (dx^i)^2\right]\nn
& & \qquad\qquad + F(r)^{\half} (H(r)\tilde{H}(r))^{\half} \left(\frac{H(r)}{\tilde{H}(r)}\right)^{\frac{3\d_1}{8}}\left(dr^2 + r^2 d\Omega_5^2\right) \nn
e^{2\phi} & = & \left(\frac{H(r)}{\tilde{H}(r)}\right)^{-3\d + \frac{7\d_1}{2}}, \qquad F_{[5]} = \frac{1}{\sqrt{2}}\left[1+\ast \right] Q {\rm Vol}(\Omega_5).
\eea
The metric is given in the Einstein frame and we have suppressed the string coupling constant $g_s$, in the above, which is 
assumed to be small. The `$*$' stands for the Hodge dual. The various functions introduced above are defined as
\bea\label{functions}
H(r) &=& 1 + \frac{\omega^4}{r^4}\nn
\tilde{H}(r) &=& 1 - \frac{\omega^4}{r^4}\nn
F(r) &=& \left(\frac{H(r)}{\tilde{H}(r)}\right)^\a \cosh^2\theta - \left(\frac{\tilde{H}(r)}{H(r)}\right)^\b \sinh^2\theta.
\eea
The solution is characterized by seven parameters $\a$, $\b$, $\d$, $\d_1$, $\theta$, $\omega$, and $Q$. However, not all of 
them are independent - rather they are constrained by the three relations
\bea\label{paramrelation}
& & \a-\b = -\frac{3}{2} \d_1\nn
& & \a + \b = \sqrt{10 - \frac{21}{2} \d^2 - \frac{49}{4} \d_1^2 + 21 \d\d_1} \equiv \gamma(\d,\d_1)\nn
& & Q = 4 \gamma \omega^4 \sinh2\theta.
\eea
The constraints allow us to eliminate three of the parameters and the `black' non-susy D3 brane solution is actually a 
four-parameter solution depending upon ($\d$, $\d_1$, $\omega$, $\theta$) . To cast the solution in a simpler form, we make a 
coordinate transformation from $r$ to $\rho$ defined by
\be\label{rtorho}
r = \rho \left(\frac{1+ \sqrt{G(\rho)}}{2}\right)^{\half}, \qquad {\rm where}, \qquad G(\rho) = 1 + \frac{4\omega^4}{\rho^4} \equiv 1 + 
\frac{\rho_0^4}{\rho^4}.
\ee
Under this coordinate transformation, the various functions introduced above look like,
\bea\label{relations}
& & H(r) = 1 + \frac{\omega^4}{r(\rho)^4} = \frac{2\sqrt{G(\rho)}}{\sqrt{G(\rho)} + 1}\nn
& & \tilde{H}(r) = 1 - \frac{\omega^4}{r(\rho)^4} = \frac{2}{\sqrt{G(\rho)} + 1}\nn     
& & \frac{H(r)}{\tilde{H}(r)} = \sqrt{G(\rho)},\,\,\,\, \left(H(r)\tilde{H}(r)\right)^{\half} dr^2 = G(\rho)^{-\frac{3}{4}} d\rho^2,\,\,\,\,
\left(H(r)\tilde{H}(r)\right)^{\half} r^2 = G(\rho)^{\frac{1}{4}} \rho^2. \nn
\eea
Plugging in \eqref{relations} into the solution \eqref{isolution} and expressing the metric in the string frame by the relation 
$ds_{\rm str}^2 = e^{\phi/2} ds^2$, we obtain the `black' non-susy D3 brane solution,
\bea\label{fsolution}
ds_{\rm str}^2 &=& F(\rho)^{-\half} G(\rho)^{-\frac{\d}{2}+\frac{\d_1}{4}}\left[-G(\rho)^{\frac{\d}{2}} dt^2 
+ \sum_{i=1}^3 (dx^i)^2\right]\nn
& & \qquad\qquad + F(\rho)^{\half} G(\rho)^{\frac{1}{4}-\frac{3\d}{8}+\frac{5\d_1}{8}}\left[\frac{d\rho^2}{G(\rho)} + \rho^2 d\Omega_5^2\right] \nn
e^{2\phi} &=& G(\rho)^{-\frac{3\d}{2} + \frac{7\d_1}{4}}, \qquad F_{[5]} = \frac{1}{\sqrt{2}}(1+\ast) Q {\rm Vol}(\Omega_5).
\eea
The function $F(\rho)$ is given as,
\be\label{frho}
F(\rho) = G(\rho)^{\a/2} \cosh^2\theta - G(\rho)^{-\b/2} \sinh^2\theta.
\ee
The metric in the original solution \eqref{isolution} has a singularity at $r=\omega$ arising from $\tilde{H}(r)$ in \eqref{functions}, 
but the coordinate change shifts the singularity to $\rho=0$ in the resultant metric \eqref{fsolution}. The parameter relations 
\eqref{paramrelation}, however, are unaffected by the change of coordinate. It can be easily checked that for the following values of the parameters
\be
\a+\b =2, \qquad \d_1 = - \frac{12}{7}, \qquad \d= -2, \quad {\rm which\,\,\, imply,} \quad \a=\frac{16}{7}, \qquad \b = -\frac{2}{7},
\ee
the solution \eqref{fsolution} reduces exactly to the standard black D3 brane solution \cite{Horowitz:1991cd} which, in the present coordinate, takes the form,
\bea\label{blackd3}
ds_{\rm blackD3}^2 &=& \tilde{F}(\rho)^{-\half} G(\rho)^{\half}\left[-G(\rho)^{-1} dt^2 
+ \sum_{i=1}^3 (dx^i)^2\right]
+ \tilde{F}(\rho)^{\half} \left(\frac{d\rho^2}{G(\rho)} + \rho^2 d\Omega_5^2\right)\nn            
e^{2\phi} &=& 1, \qquad F_{[5]} = \frac{1}{\sqrt{2}}(1+\ast) Q {\rm Vol}(\Omega_5)
\eea
where $\tilde{F}(\rho)$ is defined as,
\be
\tilde{F}(\rho) = 1 + \frac{\rho_0^4 \cosh^2\theta}{\rho^4} = F(\rho)G(\rho)^{1-\frac{\a}{2}}.
\ee 
In order to recover the BPS D3 brane solution, as usual, we have to take a double scaling limit $\rho_0 \to 0$, $\theta \to \infty$ such 
that $\rho_0^4 \cosh^2\theta \approx \rho_0^4 \sinh^2\theta \to R^4$ (fixed). In that case, $G(\rho) \to 1$ and $\tilde{F}(\rho) \to 1 + R^4/\rho^4$ 
and the solution \eqref{blackd3} reduces to the BPS D3 brane solution. The black D3 brane given in \eqref{blackd3} has a horizon at $\rho=0$. 
The temperature of the black D3 brane can be shown to have the value
\be\label{temp}
T = \frac{1}{\pi \rho_0 \cosh\theta} \longrightarrow \frac{1}{\pi\rho_0 \sinh\theta}, \quad (\rm{near\,\,\, extremality}).
\ee
It has been argued in \cite{Kim:2007qk}, that even though `black' non-susy D3 brane \eqref{fsolution} has an essential singularity at $\rho=0$, still 
it is possible to define a temperature. By comparing the solution, given in eq.(3.7) of \cite{Kim:2007qk}, with the solution \eqref{fsolution} in the 
present paper, supplemented with certain coordinate transformation, we find the temperature of the `black' non-susy D3 brane solution to have 
the form near extremality\footnote{It is not difficult to compute the ADM mass and also the charge of the `black' non-susy D3 brane
from the metric and the form-field given in \eqref{fsolution}. We get $M = \frac{\Omega_5 \rho_0^4}{2\kappa^2} \left[2(\a\cosh^2\theta 
+ \b \sinh^2\theta) + \frac{3}{2}\left(\d-\d_1\right)\right]$ and $|e| = \frac{\Omega_5 \rho_0^4}{\sqrt{2} \kappa} 2 (\a+\b) 
\cosh\theta \sinh\theta$ and taking the ratio we get, $\frac{\sqrt{2} \kappa M}{|e|} \geq 1$ for finite $\theta$. However, for large $\theta$,
which is assumed in this expression we find $\sqrt{2} \kappa M \to |e|$, indicating that in this limit the 
solution is near extremal. This also happens in the decoupling limit discussed below.}, 
\be\label{tempnonsusy}
T = \frac{(-2\d)^{\frac{1}{4}}}{\sqrt{\gamma}} \frac{1}{\pi\rho_0\sinh\theta}.
\ee
For the temperature to have a real value, we must demand that $\d$ be negative. Further, when $\d=-2$ and $\gamma = (\a+\b) = 2$, the 
`black' non-susy D3 brane reduces to ordinary black D3 brane and its temperature \eqref{tempnonsusy} reduces to that of the black D3 brane 
given in \eqref{temp}. On the other hand, when $\d=0$, the temperature vanishes - this is consistent with the fact that in this limit 
the metric becomes isotropic with ISO(1,3) symmetry and reduces to zero temperature non-susy D3 brane solution. 

Next, we discuss the decoupling limit of the `black' non-susy D3 brane \eqref{fsolution}. Decoupling limit is a low-energy limit in which 
the fundamental string length $\ell_s = \sqrt{\a'} \to 0$ and in analogy with black D3 brane, we make the following change of variables:
\be\label{decoup}
\rho = \a' u, \qquad \rho_0 \to \a' u_0, \qquad \cosh^2\theta = \frac{2L^4}{\gamma u_0^4 \a'^2}.
\ee
As we take $\a' \to 0$, the variable $u$ and the parameter $u_0$ which have dimensions of energy, are kept fixed. Also in the above 
$L^4 = 2 N g_{\rm YM}^2 = R^4/\a'^2$, similar to the BPS D3 brane \cite{Maldacena:1997re}, is kept fixed.
Now substituting \eqref{decoup}, the decoupled geometry of `black' non-susy D3 brane given in \eqref{fsolution} assumes the form
\bea\label{geometry}
ds_{\rm str}^2 &=& \a'\left[\left(\frac{\sqrt{\gamma/2}u_0^2}{L^2}\right)F(u)^{-\half} G(u)^{-\frac{\d}{2}+\frac{\d_1}{4}}\left(-G(u)^{\frac{\d}{2}} dt^2 
+ \sum_{i=1}^3 (dx^i)^2\right)\right.\nn
& & \qquad\qquad \left. +\left(\frac{L^2}{\sqrt{\gamma/2}u_0^2}\right) F(u)^{\half} G(u)^{\frac{1}{4}-\frac{3\d}{8}+\frac{5\d_1}{8}}
\left(\frac{du^2}{G(u)} + u^2 d\Omega_5^2\right)\right]\nn            
e^{2\phi} &=& g_s^2 G(u)^{-\frac{3\d}{2} + \frac{7\d_1}{4}}, \qquad {\rm with} \quad F(u) = G(u)^{\frac{\a}{2}} - G(u)^{-\frac{\b}{2}}, \quad 
G(u) = 1 + \frac{u_0^4}{u^4}
\eea
where we have restored the string coupling constant $g_s$ in the dilaton expression. To check the correctness of the decoupling limit 
and the geometry \eqref{geometry}, we notice that for $\d=-2$, $\gamma=2$, $\b=-2/7$, the metric reduces to the Schwarzschild 
black hole solution and for $u_0 \to 0$, the decoupling limit \eqref{decoup} reduces to that of BPS D3 brane and the metric reduces 
to AdS$_5$ $\times$ S$^5$ form.

Since here we are dealing with non-supersymmetric solution, one might wonder whether the solution is dynamically stable.
We already encounter such instability for black D3 brane as it is well-known that black D3 brane actually suffers from Gregory-Laflamme
instability \cite{Gregory:1993vy}. Still one can obtain the decoupling limit and the AdS/CFT correspondence in this case gets extended 
to AdS black hole
and finite temperature Yang-Mills theory. The natural question in this context would be: if the original black D3 brane solution is unstable then
how can one discuss about the gauge/gravity duality? The reason gauge/gravity duality makes sense is that the black D3 brane we 
consider are near-extremal\footnote{We would like to thank Juan Maldacena
for an e-mail correspondence on this issue.}. Since the extremal or the BPS D3 branes are perfectly stable and do not have any 
Gregory-Laflamme instability \cite{Gregory:1994tw}, the near-extremal solutions also do not suffer from such instability.
Note that the `black' non-susy D3 brane solution we are considering here becomes near-extremal in the decoupling limit (since
in this case $\theta$ becomes very large as we take $\a' \to 0$ by \eqref{decoup} and also by the discussion given in footnote 6) and
so it is possible that the solution is stable by the same argument as the near-extremal ordinary black D3 branes. However, we would
like to emphasize that to settle the issue of dynamical stability of the `black' non-susy D3 brane a careful analysis of the quasi-normal
modes is needed and we leave this for a future work.           

The dynamical stability does not necessarily imply that the solution is thermodynamically stable. For example, the Schwarzschild black
hole in asymptotically flat four space-time dimensions is dynamically stable but thermodynamically unstable. But this is not true
for the extended objects like branes. In fact, there is a correlated stability conjecture by Gubser and Mitra \cite{Gubser:2000ec}, 
which states that for
extended objects, dynamical stability necessarily implies thermodynamical stability. So, this should also apply for our `black'
non-susy D3 brane solution. To understand the relative stability of the ordinary black D3 brane and our `black' non-susy D3 brane 
we have calculated the free energy per unit D3 brane volume of our solution and found (for simplicity, we choose the parameters to take 
values $\a+\b=2$) that it is proportional to $-T^4 (2/|\d|)$, where $0 < |\d| \leq 2$. $\d=-2$ corresponds to ordinary
black D3 brane and the `black' non-susy D3 brane has $|\d| < 2$. Thus we see that the non-susy solution actually has more negative value
for its free energy and therefore more stable.         

In the next section, we use this geometry to compute the screening length and $Q{\bar Q}$ potential in a hot, windy non-supersymmetric 
plasma. We note here that the parameters $\d$ and $\d_1$ (notice that $\a$, $\b$ are given in terms of $\d$ and $\d_1$) can 
not take arbitrary values. First of all, since $\a$, $\b$ are real, therefore, we get a restriction on $\d$ and $\d_1$ from the
second relation in \eqref{paramrelation}. Also, for the supergravity description to remain valid $e^{2\phi}$ and the curvature of the string
metric in \eqref{fsolution} in units of $\a'$ must remain small. These two conditions will put further restrictions on the parameters
$\d$ and $\d_1$. We have taken these restrictions into account in our calculations in the following two sections. 

\vspace{.5cm}

\noindent{\it{3. Screening length and $Q{\bar Q}$ potential}} : Here we introduce a fundamental string as a probe in the gravity background 
we just described whose end points lie on the boundary $(\rho \to \infty$). The end points describing a $Q{\bar Q}$ pair is introduced
in this way in the boundary non-supersymmetric gauge theory. 
These are heavy quarks and suppose the $Q{\bar Q}$ pair is moving with a velocity $v$ along 
the $x^3$ direction of the boundary. We can pass on to the rest frame of the dipole in which the plasma is seen to move with a velocity $-v$ 
along the $x^3$ direction by inflicting the Lorentz boost 
\bea\label{boost}
& dt\rightarrow dt\cosh\eta-dx_3\sinh\eta\nn
& dx_3\rightarrow -dt\sinh\eta + dx_3\cosh\eta
\eea
where $\tanh \eta = v$ is the boost velocity. In terms of the boosted coordinates, the background reads
\bea\label{background}
ds^2 & = & \alpha'\left[\left(\frac{\sqrt{\gamma/2}u_0^2}{ 
L^2}\right)F(u)^{-\half}G(u)^{-\frac{\delta}{2}+\frac{\delta_1}{4}}\left(-\left(G(u)^{\frac{\delta}{2}}\cosh^2\eta-\sinh^2\eta\right) dt^2 \right. \right.\nn
&& \left. +\left(\cosh^2\eta-G(u)^{\frac{\delta}{2}}\sinh^2\eta\right) (dx^3)^2-\left(1-G(u)^{\frac{\delta}{2}}\right)\sinh2\eta dt dx^3+ (dx^1)^2+ (dx^2)^2\right) \nn
&& \left. +\left(\frac{L^2}{\sqrt{\gamma/2}u_0^2}\right)F(u)^{\half}G(u)^{{\frac{1}{4}}-{\frac{3\delta}{8}}+{\frac{5\delta_1}{8}}}\left(\frac{du^2}{G(u)}+u^2d\Omega_5^2\right)\right]
\label{boostedmetric}\\
& \equiv & \a' g_{\mu \nu}dx^{\mu}dx^{\nu} \nn
e^{2\phi} & = & g_s^2 G(u)^{-{\frac{3\delta}{2}}+{\frac{7\delta_1}{4}}}\label{5dilaton}.
\eea
Note from the $g_{tt}$ component of the metric in \eqref{background} that it has a singularity at a finite distance 
$u_c=u_0\left(\tanh^{\frac{4}{\delta}}\eta-1\right)^{-{\frac{1}{4}}}$.

In order to study the dynamics of a probe string in this gravity background, we need to compute the Nambu-Goto string world-sheet action
\be\label{5action1}
S= - \frac{1}{2\pi}\int d\sigma d\tau \sqrt{-\text{det}\left[h_{\a \b}\right]}.
\ee
Here $h_{\a \b}$ is the induced metric on the string world-sheet, i.e., 
\be 
h_{\a \b}= g_{\mu \nu}\frac{\pa x^{\mu}}{\pa \xi^{\a}}\frac{\pa x^{\nu}}{\pa \xi^{\b}}
\ee
and $\xi^{\a,\b}$ are the world-sheet coordinates, $\xi^{0}=\tau$ and $\xi^{1}=\sigma$. For evaluating the Nambu-Goto action, we need 
to fix the parametrization of the string world-sheet. We choose our coordinates along the brane directions in such a way that the 
dipole lies in the $x^1$-$x^3$ plane and makes an angle $\t$ with the $x^3$ direction while the Lorentz boost is in the $t$-$x^3$ 
plane\footnote{We refer the reader to Fig.3 of ref.\cite{Liu:2006he} for the picture of the situation described here.}. 
The parameterization of the coordinates are $\tau=t$, $x_1=\sigma$, $x_2=\text{constant}$, $x_3=x_3(\sigma)$ and $u=u(\sigma)$. 
If $\ell$ be the separation between the quark and the antiquark in the bound state, the projection of the dipole on the $x_1$ and 
$x_3$ directions are $\ell\sin\theta$ and $\ell\cos\theta$ respectively. At this stage, it is useful to introduce the following 
dimensionless coordinates
\be y=\frac{u}{u_0}, ~~~ x=u_0x^1, ~~~ z=u_0x^3.
\ee
In terms of the scaled coordinates, the boundary condition reads
\bea
y\left(x=\pm{\frac{u_0\ell}{2}}\sin\theta \right) & = & \Lambda \nn
z \left(x = \pm{\frac{u_0\ell}{2}}\sin\theta \right) & = & \pm{\frac{u_0\ell}{2}}\cos\theta \nonumber 
\eea 
where $x \in [-{\frac{u_0\ell}{2}}\sin\theta,{\frac{u_0\ell}{2}}\sin\theta]$ and we assume that the gauge theory lives at 
$y = \Lambda$ (we will take $\Lambda \to \infty$ in the end). 
With the parametrization fixed, we can now evaluate the relevant components of the ten-dimensional metric $g_{\mu \nu}$ and hence, 
the components of the world-sheet metric $h_{\a \b}$:
\bea\nonumber
h_{\tau\tau} & = & g_{tt}\,=\,-\left(\frac{\sqrt{\gamma/2}}{ 
L^2}\right) \frac{\left(G(y)^{\frac{\delta}{2}}\cosh^2\eta-\sinh^2\eta\right)}{F(y)^{\half}G(y)^{{\frac{\delta}{2}}-{\frac{\delta_1}{4}}}}\nn
h_{\sigma\sigma} & = & g_{11} + g_{33}\left(\frac{dx_3}{d\sigma}\right)^2+ g_{uu}\left(\frac{du}{d\sigma}\right)^2\nn
		 & = & g_{xx}+ g_{zz}z'^2+ g_{yy}y'^2\nn
& = & \frac{\sqrt{\g/2}}{L^2}\,\frac{G(y)^{-{\frac{\d}{2}}+{\frac{\d_1}{4}}}}{\sqrt{F(y)}}\left[1+(\cosh^2\eta- G(y)^{\frac{\d}{2}}\sinh^2\eta)z'^2 \right. \nn
&& ~~~~~~~~~~~~~~~~~~~~~ \left. +\frac{2L^4}{\g}F(y)G(y)^{-{\frac{3}{4}}+{\frac{\d}{8}}+{\frac{3\d_1}{8}}}y'^2\right]\\
h_{\tau\sigma} & = & 0\nonumber
\eea
Here we have also scaled $t$ by $u_0 t$ to make it dimensionless. The functions are now given as
\be
G(y) = 1 + y^{-4}, \qquad {\rm and} \qquad F(y) = G(y)^{\frac{\a}{2}} \cosh^2\theta - G(y)^{-\frac{\b}{2}} \sinh^2\theta
\ee
The world-sheet action, therefore, can be written as
\be\label{5action2}
S =-\frac{\mathcal{T}}{2\pi}\int_{-\frac{u_0\ell}{2}\sin\theta}^{+\frac{u_0\ell}{2}\sin\theta} dx{\cal L}(y(x))
\ee
where $\mathcal{T}$ is the temporal span and the Lagrangian $\mathcal{L}$ is
\bea\label{lagrangian}
{\cal L} &=& \left[- g_{tt}\left\{g_{xx}+ g_{zz}z'^2+ g_{yy}y'^2\right\}\right]^{\half}\nn
& = & \frac{\sqrt{\g/2}}{L^2}\,\frac{\sqrt{G(y)^{\frac{\delta}{2}}\cosh^2\eta-\sinh^2\eta}}{F(y)^{\half}G(y)^{{\frac{\delta}{2}}-{\frac{\delta_1}{4}}}}\,
\Big[1+(\cosh^2\eta-G(y)^{\frac{\d}{2}}\sinh^2\eta)z'^2\nn
&& \,~~~~~~~~~~~~~~~~~~~~~\, +\frac{2L^4}{\g}F(y)G(y)^{-{\frac{3}{4}}+{\frac{\d}{8}}+{\frac{3\d_1}{8}}}y'^2\Big]^{\half}.
\eea
A mere inspection of the Lagrangian tells us about the existence of two integrals of motion. Firstly, since $z$ does not appear explicitly 
in the Lagrangian, we have a constant of motion
\be\label{const1}
\frac{- g_{tt} g_{zz}}{\cal L} z' = p.
\ee
A second constant of motion arises from the fact that the Lagrangian does not depend explicitly upon $x$ either, implying 
\be\label{const2}
{\cal L}-z'\frac{\partial{\cal L}}{\partial z'}-y'\frac{\partial{\cal L}}{\partial y'} = \frac{- g_{tt} g_{xx}}{\cal L}=k = \text{constant}.
\ee
To find the scaled radial distance where the string profile turns around, we impose the condition $\frac{dy}{dx}=0$ which is tantamount to the constraint
\be 
k^2+ g_{xx}g^{zz}p^2+ g_{tt} g_{xx}\Big|_{y_t} = 0.
\ee
From the two constants of motion, we obtain
\bea
\frac{dz}{dx} & = & \frac{p}{k} g_{xx} g^{zz}\\
\frac{dy}{dx} & = & \frac{1}{k}\sqrt{ g_{xx} g^{yy}}\left[- g_{tt} g_{xx}-p^2 g_{xx} g^{zz}-k^2\right]^{\half}
\eea
which, upon integration, results in
\bea
\frac{u_0\ell}{2}\sin\theta &=& k\int_{y_t}^\infty\frac{1}{\sqrt{g_{xx} g^{yy}}\left[- g_{tt} g_{xx}-p^2 g_{xx} g^{zz}-k^2\right]^{\half}}dy \label{ysol}\\
\frac{u_0\ell}{2}\cos\theta &=& p\int_{y_t}^\infty \frac{g_{xx} g^{zz}}{\sqrt{g_{xx} g^{yy}}\left[- g_{tt}g_{xx}-p^2 g_{xx} g^{zz}-k^2\right]^{\half}}dy. \label{zsol}
\eea
The gravity solution has a IR cut-off at $y_c= \left(\tanh^{\frac{4}{\delta}}\eta-1\right)^{-{\frac{1}{4}}}$ as we mentioned after \eqref{5dilaton} 
and the turning 
point $y_t$  must satisfy $y_t\geq y_c$. $y_t$, in turn, is found by demanding that the terms in the denominator vanish separately and accepting 
the greater among the two possibilities. $y_t$ is found out numerically and it is indeed found to satisfy the lower bound $y_t \geq y_c$. At the 
same time, changing the integral variable from $x$ to $y$, the Nambu-Goto action in\eqref{5action2} is rewritten as
\be \label{totalenergy}
S= \frac{\mathcal{T}}{\pi}\int_{y_t}^\infty \frac{g_{tt} g_{xx}}{\sqrt{ g_{xx} g^{yy}}\left[- g_{tt} g_{xx}-p^2 g_{xx} g^{zz}-k^2\right]^{\half}}dy.
\ee
Inserting the appropriate expressions for the metric components, it is easy to figure out that the action is afflicted by a divergence. This is, 
in fact, not surprising and is typical of calculations of this sort. The reason for this divergence is not far too seek. The action, in this form, 
actually receives contribution from the interaction energy of the $Q{\bar Q}$ pair (which we seek to find out) and also the self-energies of the 
quark and the antiquark. Since we are interested in the interaction energy,  the next step is to get rid of the contribution coming from the self 
energies of the quark and the antiquark. To calculate the self energy, we use an open string hanging downwards from the boundary and whose end 
point on the boundary contains a quark/antiquark in fundamental representation. The relevant parametrization is $\tau=t,\,\sigma=u,\,x_3=x_3(u)$, 
and $x_1=x_2=\text{constant}$, which furnishes the following relations
\be
h_{\tau\tau} = g_{tt}, \qquad h_{\sigma\sigma} = g_{yy} + g_{zz}\left(\frac{dz}{dy}\right)^2, \qquad h_{\tau\sigma} = g_{tz}\left(\frac{dz}{dy}\right).
\ee 
The Nambu-Goto action now takes following form
\be
S_\text{free} = -\frac{\mathcal{T}}{\pi}\int_{y_c}^\infty \mathcal{L}_0 dy\nn
\ee
where 
\[\mathcal{L}_0=\sqrt{- g_{tt} g_{yy}+\left(g_{tz}^2- g_{tt} g_{zz}\right)\left(\frac{dz}{dy}\right)^2}=a(y)\sqrt{1+b(y)\left(\frac{dz}{dy}\right)^2}\]
with the following definitions for $a$ and $b$:
\bea
a(y) & = & \sqrt{- g_{tt} g_{yy}}=G(y)^{-{\frac{3}{8}}-{\frac{7}{16}}\d+{\frac{7}{16}}\d_1}\sqrt{G(y)^{\frac{\delta}{2}}\cosh^2\eta-\sinh^2\eta}\nn
b(y) & = & \left(\frac{\g}{2L^4}\right)F(y)^{-1}G(y)^{{\frac{3}{4}}-{\frac{\delta}{8}}-{\frac{3}{8}}\delta_1}\frac{G(y)^{\frac{\delta}{2}}+{\frac{3}{4}}
\left(1-g^{\frac{\delta}{2}}\right)^2\sinh^22\eta}{G(y)^{\frac{\delta}{2}}\cosh^2\eta-\sinh^2\eta}.
\eea
Further, we have multiplied the action of a free quark by $2$ to take into account the fact there is contribution to the diverging part 
from both the quark and the antiquark. $z$ being a cyclic coordinate, $\frac{\partial\mathcal{L}_0}{\partial z'}$ is a constant, say $k_0$ which yields
\be
\left(\frac{dz}{dy}\right)^2=\frac{\frac{2L^4k_0^2}{\g} F(y)^2G(y)^{-{\frac{3}{4}}+{\frac{9}{8}}\delta-{\frac{\delta_1}{8}}}\left(G(y)^{\frac{\delta}{2}}
\cosh^2\eta-\sinh^2\eta\right)}{ \left[G(y)^{\frac{\delta}{2}}+{\frac{3}{4}}\left(1-G^{\frac{\delta}{2}}\right)^2\sinh^22\eta\right]
\left[G(y)^{\frac{\delta}{2}}+{\frac{3}{4}}\left(1- G^{\frac{\delta}{2}}\right)^2\sinh^22\eta-k_0^2F(y)G(y)^{\delta-{\frac{\delta_1}{2}}}\right]} \nonumber
\ee
We expect a free string to extend right up to $y_c$. In the present case, actually the denominator can vanish at some $y$ greater than $y_c$ depending
on the value of $k_0$, thereby 
providing a potential turning point before the string hits $y_c$. This possibility is eliminated by constraining the value of the constant $k_0$ such 
that the numerator vanishes at the same point. In other words, we tune the value of $k_0$ in such a way that the zeros of the numerator and the denominator 
coincide and keeps $z'^2$ finite. This restriction enables us to extract the value of $k_0$ as
\be
k_0=\frac{2}{\sqrt{\tanh^{\frac{2\d-2\delta_1+2\alpha}{\delta}}\eta-\tanh^{\frac{2\d-2\delta_1-2\beta}{\delta}}\eta}}.
\ee
Finally, the action of two freely hanging strings is written as
\be\label{freeenergy}
S_{\text{free}}= -\frac{\mathcal{T}}{\pi}\int_{y_c}^\infty a(y)^2\sqrt{\frac{b(y)}{a(y)^2b(y)-k_0^2}} dy \nn
\ee
and the energy of the $Q \bar Q$ pair reads
\be\label{netenergy}
E = \frac{{\mathcal S}(\ell)}{\mathcal{T}} = \frac{S - S_\text{free}}{\mathcal{T}}.
\ee
We have thus obtained the formal expressions for the $Q{\bar Q}$ separation length $\ell$ or the dipole length in \eqref{ysol} and \eqref{zsol}.
We can square and add these relations to obtain $\ell$ in terms of the metric components and the constants of motion $k$ and $p$. Note that $\ell$ 
is not completely independent of $\theta$, the angle, the dipole makes with the boost direction, but depends on it through $k$ and $p$. We will
numerically solve these equations in the next section and show the variation of $\ell$ with the various parameters in the theory. The maximum
value of the dipole length is called the screening length, above which the dipole dissociates. The formal expression for the $Q{\bar Q}$ potential 
is obtained in \eqref{netenergy}. The numerical solution and the variations of the potential with the parameters of the theory will be shown and
discussed in the next section.   

\vspace{.5cm}

\noindent{\it{4. Numerical results}} :
\begin{figure}[ht]
\begin{center}\scalebox{.64} {\includegraphics[]{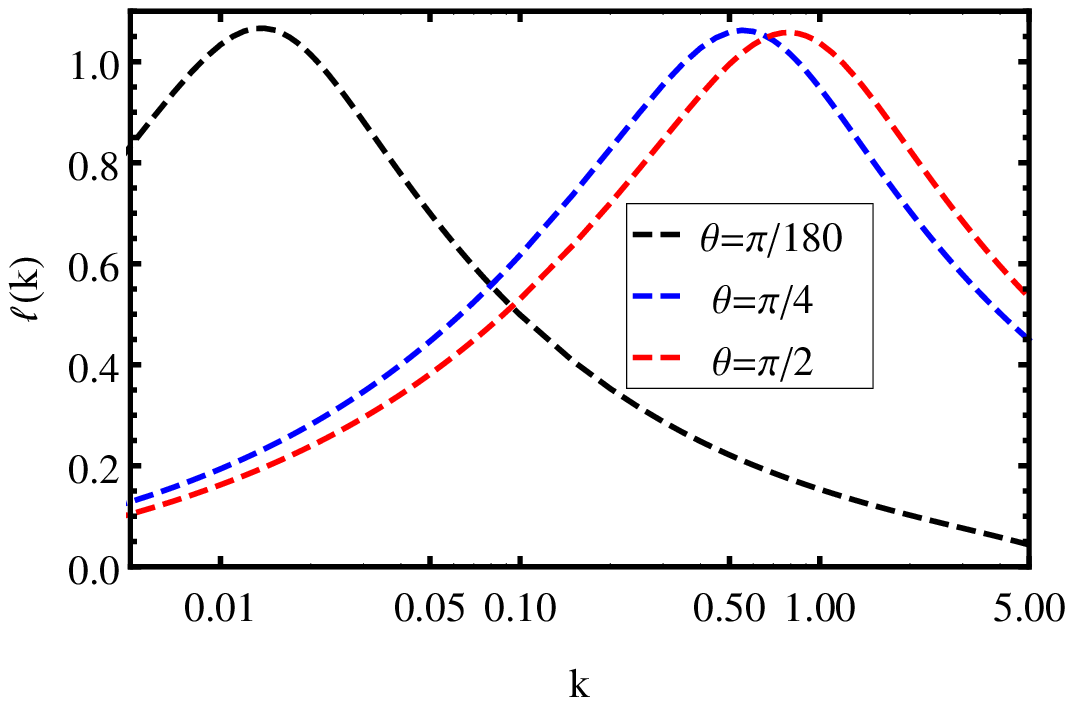}} \scalebox{.6}{\includegraphics[]{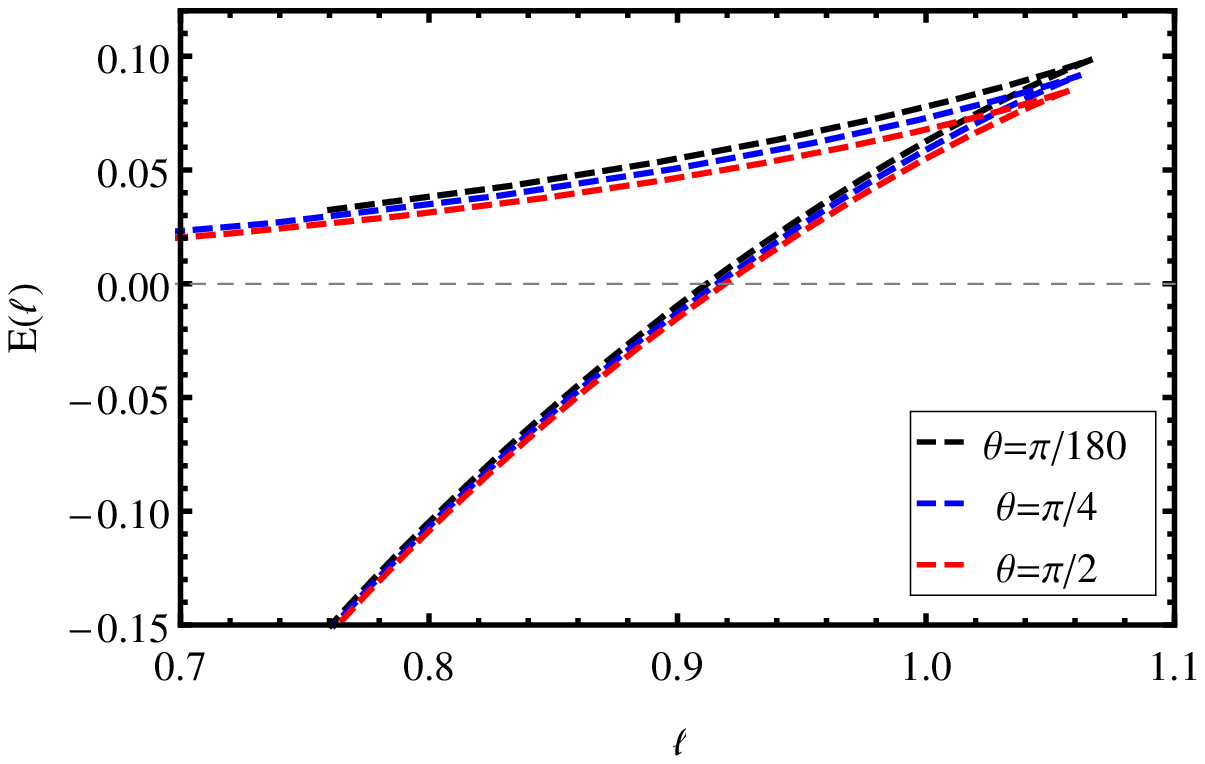}}
\caption{The plot in the left panel shows the variation of the $Q{\bar Q}$ separation length $\ell(k)$ (scaled) with $k$, a constant of motion,
given in \eqref{const2}  and the plot in the right panel shows the variation of $Q{\bar Q}$ potential $E(\ell)$ with $\ell$, when some parameters 
of the theory are fixed to the values $\delta=-0.1,\,\delta_1=-0.1$ and $\eta=1.0$. In both panels we have shown the variations for three different values
of $\theta$, the angle the dipole makes with the direction of the velocity.}\label{varytheta}
\end{center}
\end{figure}
In this section we provide the numerical results. First of all, it is important to realize that although we have found two constants of motion 
$p$ and $k$, actually they are not independent but tied through a constraint equation. This will be evident once we take the ratio of \eqref{ysol} to \eqref{zsol}. 
The L.H.S. yields $\tan \t$, whereas, the R.H.S. results in some function of $p$ and $k$. This allows us in principle to eliminate one of the constants - 
either $k$ or $p$ in 
terms of the other one and the angle $\theta$. Here, for definiteness, we shall, henceforth, choose the constant $k$ as the independent one. To evaluate the 
dipole separation ($\ell$) one needs to have knowledge of the string turning point $y_t$ that depends both upon the boost parameter $\eta$ and the integration 
constant $k$. For a given background, we have then $\ell = \ell (\eta, \theta, k)$ and $\mathcal{S} = \mathcal{S} (\eta, k)$. We can invert the first 
relationship to extract $k = k (\eta, \theta, \ell)$ and plug in into the second one to finally obtain $\mathcal{S} = \mathcal{S} (\eta, \theta, \ell)$. 
Without any loss of generality, we have set $L = u_0 = 1$ so that in our figure $\ell$ actually represents a scaled $Q\bar{Q}$ separation. 
Fig.\ref{varytheta} shows the variation of this scaled dipole length $\ell$ with the constant of motion of the string $k$ and the variation of the dipole 
potential $E$ with the dipole length $\ell$ for three different orientations of the dipole. We have set the boost parameter $\eta$ to unity while the 
two remaining parameters of the gravity background are set to $\delta = \delta_1 = -0.1$. 
The $\ell(k)-k$ plots tells us that as the orientation angle $\theta$ increases, the screening length occurs for higher values of the integration 
constant $k$. The maximum possible value of the dipole length, i.e., the screening length changes only marginally with the orientation angle $\theta$. 
The $E-\ell$ plot shows that the screening length is maximum when the dipole is almost parallel to the direction of boost velocity, 
i.e., $\theta \sim 0$\footnote{We exclude $\t=0$ because it is not allowed for the parametrization we have used for the calculation of 
$\ell$ or the potential in the previous section
and we need to use a different parametrization to include this case.}. The screening length decreases with $\theta$ and takes minimum value when dipole is exactly 
perpendicular to the direction of boost, i.e., $\theta=\pi/2$. The $E(\ell)-\ell$ plot also shows that the dipole energy is practically insensitive 
to its orientation with respect to the boost direction. 
\begin{figure}[ht]
\begin{center}\includegraphics[height=5.5cm,width=0.5\textwidth]{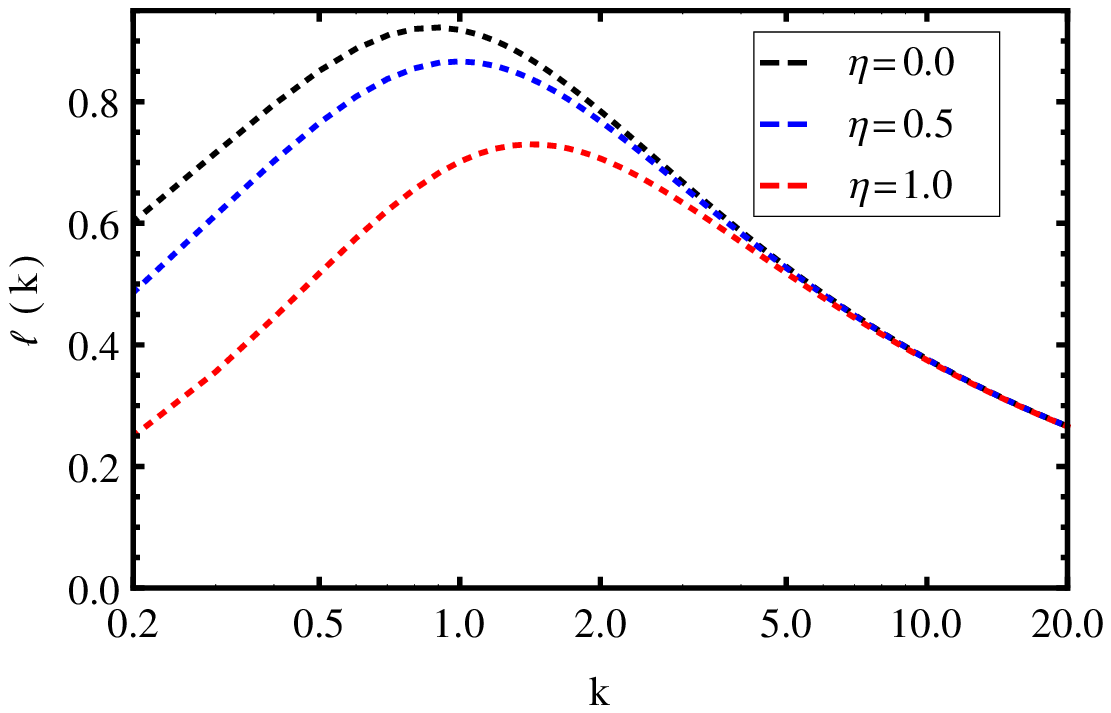}\includegraphics[height=5cm,width=0.5\textwidth]{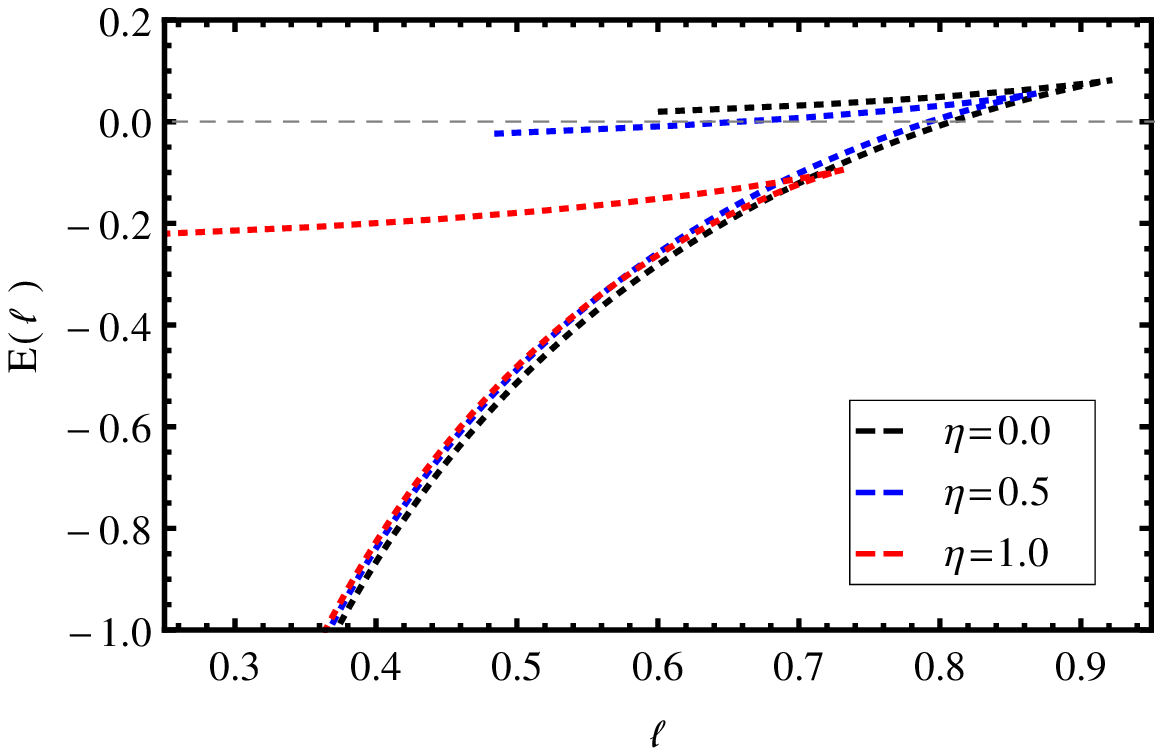}
\caption{The plot in the left panel shows the variation of the $Q{\bar Q}$ separation length $\ell(k)$ (scaled) with $k$, a constant of motion,
given in \eqref{const2}  and the plot in the right panel shows the variation of $Q{\bar Q}$ potential $E(\ell)$ with $\ell$, when some parameters 
of the theory are fixed to the values $\delta=-1.0,\,\delta_1=-1.0$ and $\theta=\pi/2$. In both panels we have shown the variations for three different 
values of $\eta$, where $\tanh \eta = v$.} \label{varyeta}
\end{center}
\end{figure}
Fig.\ref{varyeta} again shows the variation of $\ell$ with $k$ and the variation of $E$ with $\ell$ for the values of $\delta = \delta_1 = -1.0$, 
but this time we have fixed the value of $\theta$  at $\pi/2$. Instead, we have shown the variations for three different values of the boost parameter 
$\eta$. Again, the interaction energy is not much affected by the presence of a velocity. In this context, it is worth mentioning that we are only 
concerned with the lower branch of the $E-\ell$ curves since they represent the lower energy states and are more stable compared to the upper branches 
which have higher energy and does not represent stable configuration of the dipole. However, the $\ell(k)-k$ plot clearly shows that the screening 
length is quite sensitive to the boost parameter and is the maximum when $\eta =0$. This means the dipole is most stable (for this particular 
configuration) when it is sitting still in the plasma - any motion through the hot plasma leads to a decrease in the screening length and makes 
the dipole more vulnerable to dissociation. This is expected because as the velocity increases there is more chance of a collision of the dipole
with the background making it more easily dissociable. 
\begin{figure}[ht]
\begin{center}\includegraphics[height=5.5cm,width=0.5\textwidth]{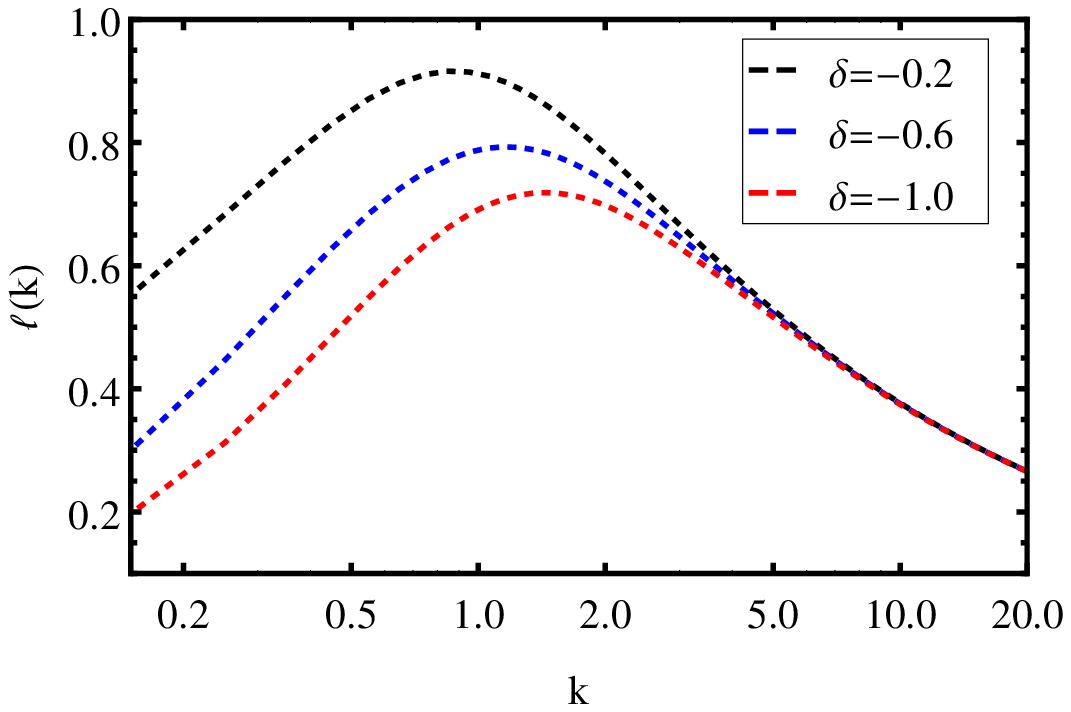}\includegraphics[height=5cm,width=0.5\textwidth]{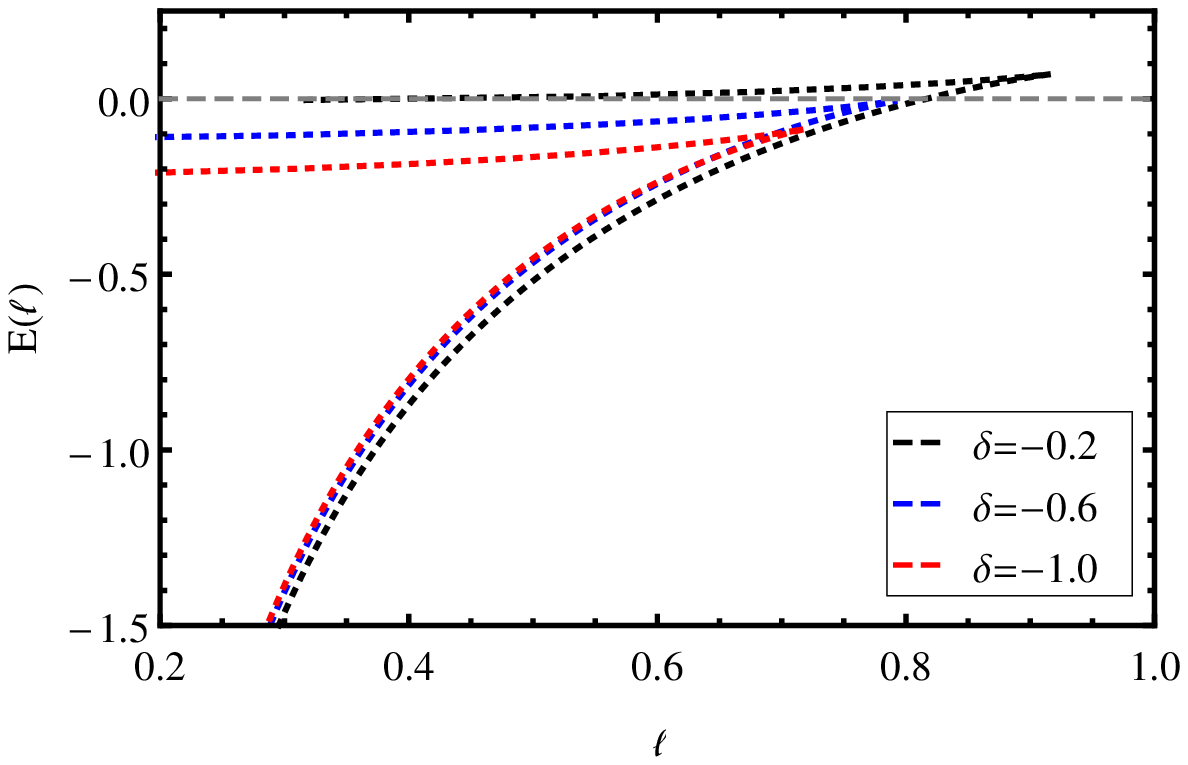}
\caption{The plot in the left panel shows the variation of the $Q{\bar Q}$ separation length $\ell(k)$ (scaled) with $k$, a constant of motion,
given in \eqref{const2}  and the plot in the right panel shows the variation of $Q{\bar Q}$ potential $E(\ell)$ with $\ell$, when some parameters 
are fixed to the values $\eta=1.0,\,\theta=\pi/2$ and $\theta=\pi/2$, $-(3/2)\delta+(7/4)\delta_1=-0.5$. Keeping the the last combination fixed
keeps the effective coupling of the theory fixed. In both panels we have shown the variations for three different values of $\d$ which is directly related
to the temperature.} 
\label{varydelta}
\end{center}
\end{figure}

\begin{figure}[ht]
\begin{center}\includegraphics[height=5.5cm,width=0.5\textwidth]{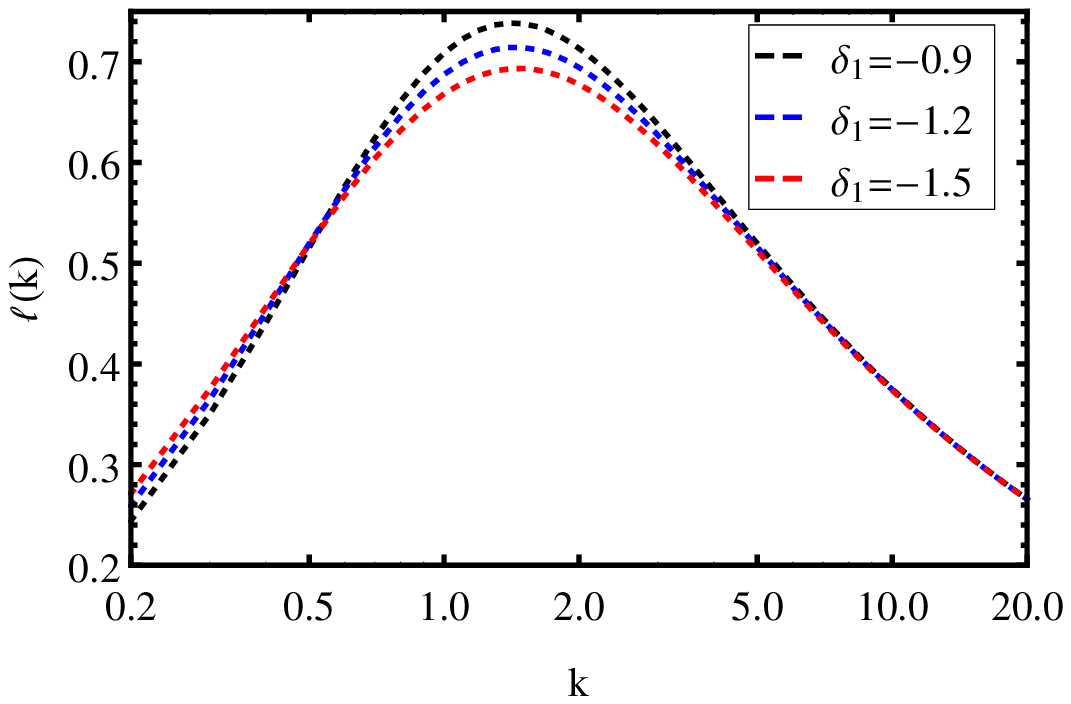}\includegraphics[height=5cm,width=0.5\textwidth]{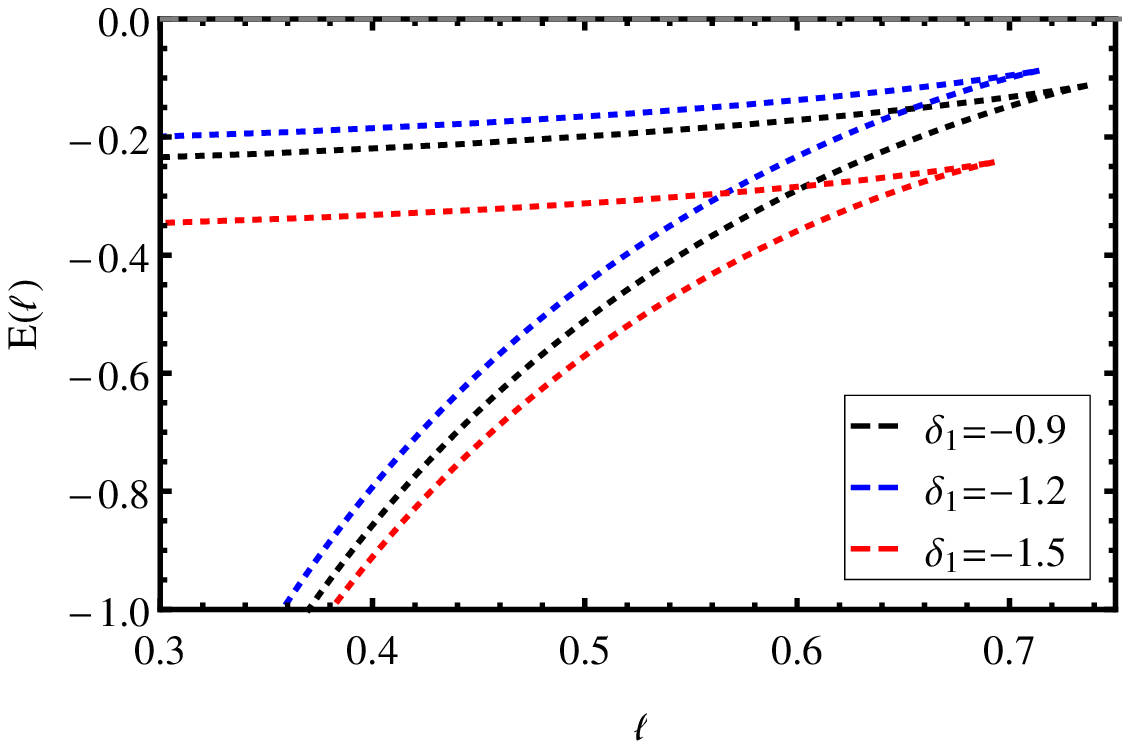}
\caption{The plot in the left panel shows the variation of the $Q{\bar Q}$ separation length $\ell(k)$ (scaled) with $k$, a constant of motion,
given in \eqref{const2}  and the plot in the right panel shows the variation of $Q{\bar Q}$ potential $E(\ell)$ with $\ell$, when some parameters 
are fixed to the values $\eta=1.0,\, \delta=-1.0$, and $\theta=\pi/2$. In both panels we have shown the variations for seven different values of 
$\d_1$ which is directly related to the effective coupling of the theory.}\label{varydelta1}
\end{center}
\end{figure}

The next two figures Fig.\ref{varydelta} and Fig.\ref{varydelta1} show the variations $\ell$ with $k$ and also 
the variations of $Q{\bar Q}$ potential $E$ with $\ell$ when some parameters of the gravity theory are varied. The parameters which we
denoted $\d$ and $\d_1$ in the non-susy D3 brane configuration \eqref{fsolution} do not appear in the supersymmetric theory
and therefore the plots under their variations reveal new features not observed before (can not be compared with the
supersymmetric theory). We also give some physical interpretations of the parameters.
Fig.\ref{varydelta} shows the behavior of the curves $\ell$ with $k$ and also $E(\ell)$ with $\ell$ when the parameter $\d$ is varied.
Here we have kept $\theta$, $\eta$ and the combination $-\frac{3}{2}\d + \frac{7}{4}\delta_1$ fixed to the values $1.0$, $\pi/2$ 
and $-0.5$ respectively. The reason for keeping the particular combination of $\d$ and $\d_1$ fixed
is that this combination appears in the expression for the dilaton given in \eqref{geometry} and so, keeping this combination fixed will
keep the effective coupling, $g_{\text{eff}}$ = fixed. We find that the screening length 
of the dipole have a strong dependence on the value of $\delta$, while the interaction energy changes mildly. As $|\delta|$ increases, the screening length 
decreases. It is not difficult to understand why this is so, if we remember that non-susy D3 brane has a temperature near extremality
proportional to $\d$ (given in \eqref{tempnonsusy}). Expressed in terms of $u_0$ it has the form,
\be\label{5temp}
T=\frac{(-2\delta)^{1/4}}{\sqrt{2}\pi L^2}u_0=\frac{(2|\delta|)^{1/4}}{\sqrt{2}\pi L^2}u_0, \qquad {\rm as,} \quad \delta \leq 0
\ee
Therefore as $|\d|$ increases the temperature of the system increases which makes the dipole to dissociate more easily resulting
in the decrease of the screening length. Also as the temperature increases the interaction energy of the $Q{\bar Q}$ decreases making
the dipole less stable and that is seen in the figure as the slight increment of $E(\ell)$ with $|\d|$. 
Next, Fig.\ref{varydelta1} explores how the $\ell-k$ plot and the $E-\ell$ plot depend upon $\delta_1$ when all other 
parameters are kept fixed. Now the screening length changes mildly with variation in $\delta_1$ while the corresponding change in the potential 
energy is more prominent. Let us now see how the parameter $\delta_1$ affects the screening length and the potential. We have already 
seen that the temperature of the non-susy brane (which is also interpreted as the temperature of the gauge theory) depends upon $\delta$ for 
fixed value of $u_0$, whereas, the effective coupling $g_{\text{eff}}$ (related to the dilaton field) depends upon both $\delta$ and $\delta_1$. 
Now if we vary $\delta_1$, keeping $\d$ fixed, the temperature remains fixed but the effective coupling changes. 
From \eqref{5dilaton}, we find the dimensionless gauge coupling to be
\be \label{5coupling}
g^2_\text{eff} \sim Ne^\phi=Ng_sG(u)^{{\frac{3}{2}}|\delta|-{\frac{7}{4}}|\delta_1|},\qquad \text{as} \qquad \delta,\,\delta_1\le0
\ee
At a fixed energy scale $u$, the coupling $g_\text{eff}$ decreases with the increasing $|\delta_1|$. As the coupling becomes weaker, 
the quark-antiquark pair is loosely bound and the maximum allowed length of the dipole decreases - this is clearly visible in Fig.\ref{varydelta1}.
On the other hand, as the coupling becomes weaker with increasing $|\d_1|$, the dipole will be loosely bound and becomes less stable. This is
actually seen in the $E(\ell)$ vs. $\ell$ plot in Fig.\ref{varydelta1}. Indeed we see that the physically relevant lower portion of the
curve actually goes up as we increase $|\d_1|$. But this happens only upto certain value of $|\d_1|$ between 1.3 and 1.4. But beyond that
the effect gets reversed. As we increase $|\d_1|$ further the potential goes down showing that the dipole becomes more stable. This latter 
stability of the dipole is quite counterintuitive and we do not have a satisfactory physical explanation for its occurrence.

\begin{figure}[ht]
\begin{center}\includegraphics[height=5cm,width=0.5\textwidth]{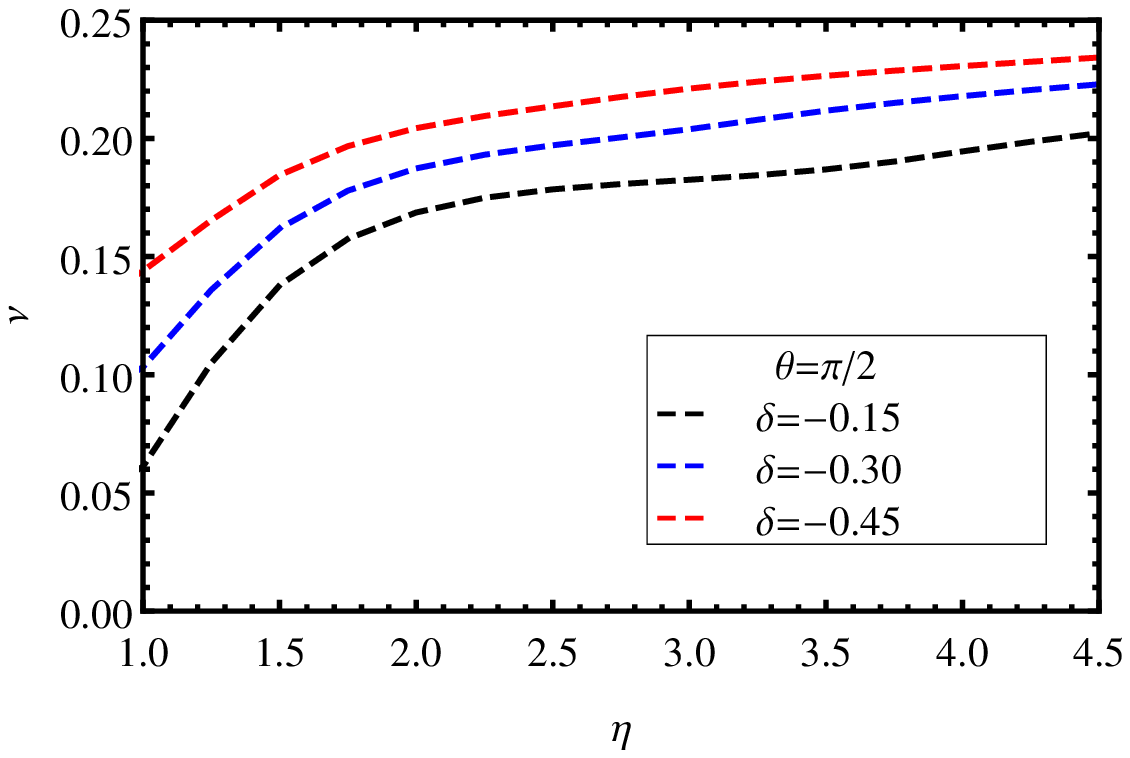}\includegraphics[height=5cm,width=0.5\textwidth]{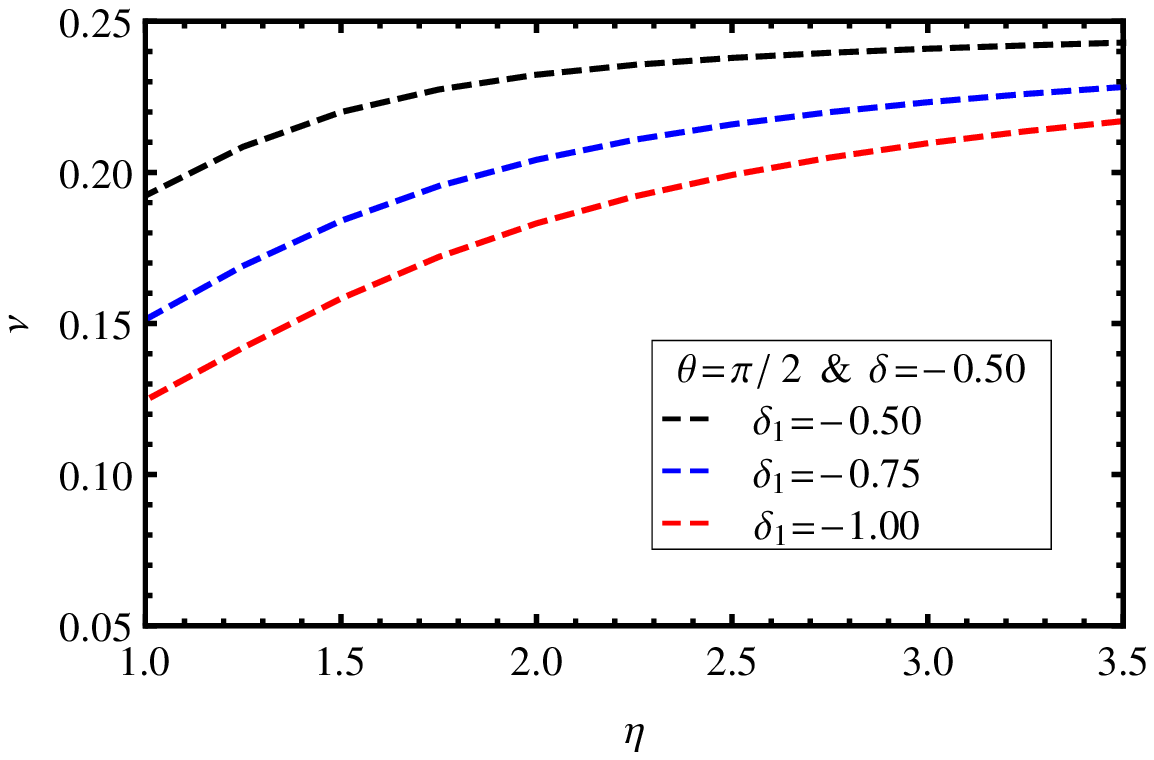}
\caption{Here in both the panels we have shown the variations of the exponent $\nu$ (defined in eq.\eqref{screen}) with $\eta$
(i.e., the velocity). In the left panel the parameters are fixed to the values $\theta=\pi/2$ and $-(3/2)\delta+(7/4)\delta_1=-0.5$ (this means the effective
coupling is fixed) and in the right panel the parameters are fixed to the values $\theta=\pi/2$ and  $\delta=-0.5$ (this means the temperature is fixed).} 
\label{lmax}
\end{center}
\end{figure}     
Lastly, we mention that for the superconformal theory as in $D=4$, ${\cal N} = 4$ SYM theory, it is known that the variation of the
screening length with velocity is given as \cite{Liu:2006he},
\be\label{screen}
\ell_\text{max}(v) \sim \left(1-v(\eta)^2\right)^{\nu}\ell_{\rm max}(0)
\ee
where $\ell_{\rm max}(0)$ is the screening length when the background is at rest and $\nu$ takes a value $0.25$. However, it is also observed
in \cite{Liu:2006he}, that for non-conformal theory the power $\nu$ in \eqref{screen} actually drops from the value $0.25$. Since in
our background there is no conformal symmetry, we would like to see how the exponent changes with velocity. This can be 
obtained by studying the equations \eqref{ysol} and \eqref{zsol}. In Fig.\ref{lmax} we have shown this variation. In the first figure we 
have shown the variation of $\nu$ with $\delta$ (i.e., the temperature) while keeping $\theta$ and $-\frac{3}{2} \d + \frac{7}{4} \d_1$ (i.e., 
effective coupling) fixed and in the second figure we have shown the variation of $\nu$ with $\d_1$ (i.e., the effective coupling) while
keeping $\theta$ and $\d$ (i.e., temperature) fixed. For high velocity, we see that in both cases,
the curves saturate to some value less than $0.25$, a characteristic of a non-conformal background.

\vspace{.5cm}

\noindent{\it{5. Conclusion}} : To conclude, in this paper we have shown an application of non-supersymmetric AdS/CFT in the calculation
of Wilson loop in QGP. The non-supersymmetric AdS/CFT has been proposed earlier by two of us, by proposing a decoupling limit for the
non-supersymmetric D3 brane solution of type IIB string theory. By this procedure we obtained the throat geometry of the non-susy D3 brane
and that gave us the gravity dual of a non-supersymmetric gauge theory on the boundary. We have taken a `black' version of this solution and the
decoupled geometry of this solution has given a finite temperature, non-supersymmetric, non-conformal gauge theory on the boundary. We have taken
this geometry and boosted it along one of the brane directions. We have also introduced a probe string in this background whose end points
represent the quark-antiquark pair on the boundary theory. The $Q{\bar Q}$ pair has been made to lie at an angle $\theta$ from the direction
of the boost.
Then following LRW \cite{Liu:2006he} we have computed the time-like Wilson loop in this background which in turn has given us the formal
expression of $Q{\bar Q}$ separation length \eqref{ysol}, \eqref{zsol} and the interaction potential \eqref{netenergy}. We have numerically solved
those equations and plotted the $Q{\bar Q}$ separation length ($\ell$) as a function of certain integral of motion ($k$) and also the interaction energy 
($E$) of the diople as a function of $\ell$. The maximum allowed $Q{\bar Q}$ separation is called the screening length since beyond this $Q$ and ${\bar Q}$
in the dipole dissociates. As happens in real heavy ion collision experiment, we have shown the variation of screening length $\ell_{\rm max}$ as well
as the interaction potential $E$ with the angle $\theta$ (angle between the dipole and the direction of the velocity), $\eta = \tanh^{-1} v$ ($v$, the 
background velocity) and also with $|\d|$ (related to the temperature of the gauge theory) and $|\d_1|$ (related to the effective gauge coupling),
keeping other parameters fixed. 
These variations with $\theta$ and the velocity have also been studied for supersymmetric theory by Liu, Rajagopal and Wiedemann \cite{Liu:2006he} 
and surprisingly, 
we found qualitative agreement with their results. This clearly shows that these results are quite robust as they do not depend on the presence of any
supersymmetry (also the conformal symmetry) of the theory. We have also shown how the screening length and the potential change with the change
of $|\d|$ (i.e., temperature) and $|\d_1|$ (i.e., the effective coupling) not seen in the supersymmetric theory. We observed some peculiar
behavior of the potential when the coupling changes. Initially when $|\d_1|$ increases, the potential slightly increases as expected, but beyond
some value in between 1.3 and 1.4, the effect gets reversed, namely, as $|\d_1|$ increases further the potential suddenly decreases, whose
physical explanation is not clear to us.

One of the motivations for studying QGP behavior using AdS/CFT for LRW \cite{Liu:2006ug} was to see the observed quarkonium suppression
when the background is not static but has a velocity. They calculated the screening length for the superconformal theory and found that the
screening length gets reduced with velocity by a factor $(1-v^2)^{1/4}$ from its static value. This indeed supports the observed quarkonium
suppression as the QGP produced in heavy ion collision moves with high relative velocity with the dipole. They also observed that for theories 
without conformal symmetry
the exponent $\nu$ should have values less than $0.25$. Since the non-supersymmetric theory we are dealing with also does not have a conformal
symmetry we have plotted the exponent $\nu$ with velocity. We found that at high velocity, $\nu$ indeed saturates to values somewhat below
$0.25$, when both $\d$ and $\d_1$ are varied confirming that for theories without conformal symmetry, the velocity dependence of screening length
would be such that it would enhance the quarkonium suppression.

We remark that this work is probably the first attempt to study QGP properties using non-supersymmetric AdS/CFT following from the decoupling limit 
of non-susy D3 brane. There are various other properties of QGP, like jet quenching parameter, thermalization, phase transition, chiral symmetry breaking, 
photon and dilepton production etc. which can also be studied using the background described in this paper. We hope to come back to some of these 
issues in future.   

\vspace{1cm}

\noindent{{\it Acknowledgement}} : One of us (SR) would like to thank Juan Maldacena for an e-mail correspondence.

\end{document}